# Stimulated emission of surface plasmon polaritons

M. A. Noginov [1], G. Zhu [1], M. F. Mayy [1], B. A. Ritzo [1], N. Noginova [1], and V. A. Podolskiy [2]
[1)] Center for Materials Research, Norfolk State University, Norfolk, VA 23504
[2)] Department of Physics, Oregon State University, Corvallis, OR 9733

**Abstract**
We have observed laser-like emission of surface plasmon polaritons (SPPs) decoupled to the glass prism in an attenuated total reflection setup. SPPs were excited by optically pumped molecules in a polymeric film deposited on the top of the silver film. Stimulated emission was characterized by a distinct threshold in the input-output dependence and narrowing of the emission spectrum. The observed stimulated emission and corresponding to it compensation of the metallic absorption loss by gain enables many applications of metamaterials and nanoplasmonic devices.

Photonic metamaterials, engineered composites with unique electromagnetic properties, have become in recent years a hot research topic because of their interesting physics and exciting potential applications, which include but are not limited to imaging with sub-diffraction resolution [1-5], optical cloaking [6,7] and nanolasers [8-10]. Localized surface plasmons (SPs) and propagating surface plasmon polaritons (SPPs) play a key role in design of majority of metamaterials. Unfortunately, in contrast to mid- and far-IR structures [11,12], the performance of optical SP- and SPP-based systems is fundamentally limited by absorption loss in metal, which causes a reduction of the quality factor of SPs and shortening of the propagation length of SPPs. The solution to the loss problem, which has been proposed over the years in a number of publications [13-16], is to add an optical gain to a dielectric adjacent to the metal. A substantial enhancement by optical gain of localized SPs in aggregated silver nanoparticles, evidenced by six-fold increase of the Rayleigh scattering, was demonstrated in [17,18]. A principle possibility to offset SPP loss by gain has been experimentally shown in Ref. [19]; and the optical gain ~420 cm$^{-1}$ sufficient to compensate ~30% of the SPP internal loss in a silver film was achieved in Ref. [20]. **Here we report experimental observation of the stimulated emission of SPPs at optical frequency. Our work proves that the compensation of SPP loss by gain is, indeed, possible, opening the road for many practical applications of nanoplasmonics and metamaterials. Besides resolving one of the fundamental limitations of modern nanoplasmonics, the observed phenomenon adds a new emission source to the 'toolbox' of active optical metamaterials.** The obtained SPP stimulated emission is relevant to the theoretically proposed SPASER [21], the device analogous to laser designed to generate SPs. However, the existence of any feedback mechanism in our system remains questionable.

SPPs are electromagnetic waves coupled to oscillations of free-electron plasma, which propagate at the interface between metal, characterized by the dielectric constant $\varepsilon_1$, and adjacent to it dielectric, characterized by the dielectric constant $\varepsilon_2$, Fig. 1a. SPPs are confined to the proximity of metal-dielectric interface and decay exponentially to both media. The wave vector of the SPP propagating in the $x$ direction is given by the expression $k_{SPP} = \frac{\omega}{c}\sqrt{\frac{\varepsilon_1 \cdot \varepsilon_2}{\varepsilon_1 + \varepsilon_2}}$, where $\omega$ is the frequency and c is the speed of light [22]. SPPs can be excited, for example, by a $p$ polarized light incident onto metallic film from the side of a glass prism at the critical angle $\theta_0$, at which the projection of the wave vector of phonon to the $x$ axis, $k_x^{phot} = \frac{\omega}{c} n_0 \sin\theta_0$, is equal to $k_{SPP}$, Fig. 1a. Once excited, SPPs propagate alone the metallic surface, simultaneously decoupling to the prism at the same resonant angle $\theta_0$.

In this work, 39 nm - 81 nm silver films were deposited on the glass prism with the index of refraction $n_0 = \sqrt{\varepsilon_0} = 1.7835$ and coated with 1 μm - 10 μm polymethyl methacrylate (PMMA) films doped with rhodamine 6G (R6G) dye in concentration $2.2 \times 10^{-2}$M, Fig. 1a. The same dye-doped polymer was used as a gain medium of random laser in Ref. [23].

Four different sets of experiments have been performed.

(1) In the first set of measurements, used primarily for calibration purposes, we excited SPPs from the bottom of the prism and measured the reflectivity of the sample $R$ as a function of the incidence angle $\theta$, Fig. 1b. At the resonance angle $\theta_0$, the energy of incident light was transferred to SPPs, yielding a minimum in the reflectivity profile $R(\theta)$ [22]. This standard experiment accompanied by theoretical modeling [22,20] allowed us to determine the dielectric constants of particular silver films studied. The width of the reflectivity profile $R(\theta)$ corresponds to the propagation length of SPP, $L = [2(\gamma_i + \gamma_r)]^{-1}$, with $\gamma_i$ and $\gamma_r$ being, respectively, intrinsic losses due to material absorption and radiation losses due to SPP decoupling to the prism.

In the remaining experiments, the samples were excited from the PMMA side at nearly normal angle of incidence.

(2) Unintentional scatterers are always present in polymeric films and can even provide for a feedback in polymer random lasers [24]. In our system scatterers enable excitation of surface plasmon polaritons *via* coupling of incident light into SPP modes. Experimentally, SPPs were excited *via* pumping the PMMA film, and the intensity of light $I$ decoupled to the prism was studied as a function of angle $\theta$. The angular profile of the decoupled light intensity $I(\theta)$ had a maximum almost at the same resonance angle $\theta_0$, at which the reflectivity profile $R(\theta)$ had its minimum, Fig. 1b. This confirms that the detected light was, indeed, due to pumping-induced SPPs. The demonstrated effortless method of excitation of SPPs does not require any prism or grating and can be regarded as a 'poor man' technique.

(3) Alternatively, the SPPs can be excited *via* emission of R6G molecules. The laser light at $\lambda$=532 nm corresponds to the absorption maximum of R6G. It excites dye molecules in the PMMA/R6G volume, in particular, in the vicinity of the silver film where the pumping-induced SPP is confined. Excited dye molecules, in turn, partly emit to SPP modes at corresponding frequencies. The SPPs excited by optically pumped molecules (reported earlier in Refs. [25,26]) get decoupled to the prism at the angles corresponding to the SPPs' wavenumbers. At low pumping intensity, the SPP spectra recorded at different angles $\theta$ qualitatively resembled the R6G spontaneous emission spectrum modulated by the SPP decoupling function, Fig. 1c Furthermore, the angular emission profiles $I(\theta)$ recorded at different wavelengths matched the reflectivity profiles $R(\theta)$, Fig. 1d. This confirms that the observed emission was, in fact, due to decoupling of SPPs excited by pumped R6G molecules.

(4) The character of SPP emission excited *via* optically pumped dye molecules has changed dramatically at high pumping intensity.

(*i*) The emission spectra considerably narrowed in comparison to those at low pumping, Fig.2a.

(*ii*) The narrowed emission spectra peaking at ~602 nm became almost independent of the observation angle.

(*iii*) The dependence of the emission intensity (recorded in the maximum of the emission spectrum) on the pumping intensity was strongly nonlinear with the distinct threshold, Fig. 2b.
(*iv*) The value of the threshold $I_{th}$ depended on the observation angle $\theta$. The angular dependence $I_{th}(\theta)$, resembled the angular profile of the reflectivity $R(\theta)$, Fig. 2c.

The experimental results above suggest the *stimulated* character of emission decoupled from SPP mode(s). To prove that the stimulated emission seen in our experiments comes from direct generation of SPPs rather than from random lasing in PMMA/R6G film decoupled to prism *via* SPPs, we studied the emission collected from the back side of the prism as well as from glass slides with similar deposited films, close to the normal angle of incidence and at a grazing angle. At high pumping energy, the collected emission spectrally narrowed and demonstrated a threshold input-output behavior. However, the values of the thresholds (higher at nearly normal direction than at a grazing angle) were significantly larger than the threshold in the case of SPPs decoupled to the prism, and the narrowed emission spectra had their maxima at wavelengths as short as ~584 nm. We, thus, conclude that the emission decoupled to the glass prism was generated by SPPs rather than originated in purely photonic modes within the PMMA/R6G layer above the silver film.

To further analyze the SPP emission observed in our experiments, we numerically solve Maxwell's equations with transfer matrix approach. In our simulations we first approximate the emitter by a point dipole and represent its field as a series of plane waves with well-defined *x* components of the wave vectors through Fourier expansion. We then solve for propagation of individual components of this expansion using the transfer-matrix method (TMM) [27], and finally represent the total field resulting from coupled to SPP emission as a sum of all single-component solutions. This process allows us to directly calculate the field distribution in the system as well as to analyze angular and spectral distributions of the emitted signals below lasing threshold. The excellent agreement between experimental and theoretical SPP emission spectra at three different decoupling angles is shown in Fig. 1c.

The TMM simulations, assuming linear behavior of the system in the frequency domain, are inapplicable above the lasing threshold. To qualitatively assess the onset of lasing oscillations, we use the following Fourier-transform-based approach. We first assume that SPP propagating in the *x* direction and decaying along its propagation is characterized by the complex wave vector $k_x = k_x' + ik_x''$, such that

$$E = E_0 e^{ik_x' x} e^{-k_x'' x} \tag{1}$$

where $E$ is the SPP electric field and $E_0$ is its amplitude at $x=0$. The spectrum of this SPP in a wavenumber domain is given by the Fourier transform

$$E(k) = \frac{1}{\pi} \int_0^{x'} E(x) e^{-ikx} dx = \frac{1}{\pi\left[-k_x'' + i\left(k_x' - k\right)\right]}, \tag{2}$$

where $x'=\infty$. It corresponds to Lorentzian distribution of decoupled light intensity about the resonant value of the wave vector and, in particular, determines the shape of the dip in the reflectivity profile $R(\theta)$, which is routinely recorded in SPP experiments in the attenuated total reflection geometry, Fig. 1a.

When gain in the system overcomes loss, the negative sign in the second exponent of Eq. (1) changes to positive, which corresponds to exponential increase of the SPP intensity. In this case, if the upper limit of integration in Eq. (2) is set to be equal to infinity, the integral diverges, which physically corresponds to the fact that no steady-state gain can exist (or be physically supported by any source of pumping) in an infinitely large system. However, if the upper limit of integration is set at a finite value (the condition, which simulates a pumped spot of a finite size) then the integral can be readily solved analytically.

The Lorentzian-like spectrum of SPP emission $I_{SPP}(k) \propto |E(k)|^2$, calculated for SPP propagation with amplification, has a maximum at the resonance wavenumber $k_x'$, Fig. 3a, in agreement with the experimentally observed reduction of the SPP lasing threshold at the resonant angle, Fig. 2c. When emission intensities $I_{SPP}(k)$ calculated at fixed wave vector $k_x'$ and different values of gain $k_x''$ (pumping energies) are plotted against $k_x''$, the dependence, characterized by a distinct threshold, has a qualitative similarity with input-output curves measured experimentally, Figs. 2b and 3b. Knowing the concentration of R6G molecules in the film ($1.3 \times 10^{19}$ cm$^{-3}$), absorption and emission cross sections of R6G ($\sigma_{abs}^{\lambda=532nm}=4.3 \times 10^{-16}$ cm$^2$, $\sigma_{em}^{\lambda=600nm}=1.9 \times 10^{-16}$ cm$^2$), and life-time of excited R6G molecules at given concentration of dye, $\tau=1$ ns [28], we were able to evaluate the experimental threshold gains to be of the same order of magnitude as the theoretically predicted ones satisfying the condition $(\gamma_i+\gamma_r)=0$. Thus, although nonlinear dependence of the SPP intensity on the pumping intensity and narrowing of the emission spectrum are expected even below the threshold, a good agreement between the theoretical and the experimental results serves as a strong indication that the observed emission is due to stimulated emission of SPPs in a regime when the total gain in the system exceeds the total loss.

To summarize, we have studied surface plasmon polaritons (SPPs) excited by emission of optically pumped rhodamine 6G (R6G) molecules and by direct scattering of pumping light in a polymeric film in the attenuated total reflection setup. We have observed laser-like emission of SPPs decoupled to the glass prism. In particular, the input-output dependence had a distinctive threshold and the emission spectrum narrowed considerably above the threshold. The threshold and the overall emission intensity strongly depended on the SPP decoupling angle. Thus, the threshold was minimal and the emission intensity was maximal at the resonance condition, at which the wavevector of the SPP $k_{SPP}$ was equal to the projection of the photon wavevector to the metal-dielectric interface $k_x^{phot}$. This behavior is in a good agreement with the model considering SPP propagation with amplification, when the total gain in the system exceeds the total loss. Random laser effect, which occurred in R6G doped polymeric film at high pumping, had different spectrum and different threshold than the SPP emission. That is why we ruled it out as a possible source of the laser-like SPP behavior. The experimental thresholds of SPP stimulated emission had the same order of magnitude as the theoretically predicted ones. The observed phenomenon may be relevant to SPASER proposed in Ref. [21]. However, the existence of the stimulated emission feedback, which is a key element of SPASER, was not confirmed in our system. The demonstrated phenomenon adds a new stimulated emission source to the toolbox of nanophotonic materials and devices and proves that total compensation by metamaterials' loss by gain is, indeed, possible.

The work was supported by the NSF PREM grant # DMR 0611430, the NSF CREST grant # HRD 0317722, the NSF NCN grant # EEC-0228390, the NSF grant # ECCS-0724763, the NASA URC grant # NCC3-1035, ONR grant #N00014-07-1-0457, the Petroleum Research Fund, and US Army. The authors cordially thank F. Javier Garcia de Abajo for useful discussions.

1. J. B. Pendry, Phys. Rev. Lett. **85**, 3966-3969 (2000).
2. Z. Jacob, L. V. Alekseyev, E. Narimanov, Opt. Express **14**, 8247-8256 (2006).
3. A. Salandrino and N. Engheta, Phys. Rev. B **74**, 075103 (2006).
4. Z. Liu, et al., Science **315**, 1686 (2007).
5. I. I. Smolyaninov, Y.-J. Hung, C. C. Davis, Science **315**, 1699-701 (2007).
6. D. Schurig, et al., Science **314**, 977 (2006).
7. W. Cai, et al., *Nature Photonics* **1**, 224 (2007).
8. P. Muhlschlegel, et. al., *Science* **308**, 1607-1609 (2005).
9. J. A. Gordon, R. W. Ziolkowski, *Optics Express* **15**, 2622-2653 (2007).
10. S. Noda, *Science* **314**, 260 (2006).
11. C. Sirtori, et al., *Opt. Lett.* **23,** 1366-1368 (1998).
12. A. Tredicucci, et al., Applied Physics Letters **76**???, 2164-2166 (2000).
13. A. N. Sudarkin, P. A. Demkovich, *Sov. Phys. Tech. Phys.* **34**, 764-766 (1989).
14. I. Avrutsky, *Phys. Rev. B* **70**, 155416 (2004).
15. M. P. Nezhad, K. Tetz, Y. Fainman, *Optics Express* **12**, 4072-4079 (2004).
16. N. M. Lawandy, *Appl. Phys. Lett.* **85**, 5040-5042 (2004).
17. M. A. Noginov, et al., Opt. Lett., **31**, pp. 3022-3024 (2006).
18. M. A. Noginov, et al., Appl. Phys. B **86**, pp. 455-460 (2007).
19. J. Seidel, S. Grafstroem, L. Eng, *Phys. Rev. Lett.* **94**, 177401 (2005).
20. M. A. Noginov, et al., Proceedings of SPIE Volume: **6642**, 2007; arXiv:0704.1513 (April 2007).
21. D. Bergman, M. Stockman, *Phys. Rev. Lett.* **90,** 027402 (2003).
22. H. Raether, S*urface plasmons on smooth and rough surfaces and on gratings*, Springer-Verlag, (Berlin, 1988).
23. Y. Ling et al., Phys. Rev. A., **64**, 063808 (2001) 8 pages.
24. M. A. Noginov, *Solid-State Random Lasers*, Springer, printed in the USA, 235 p. (2005).
25. W. H. Weber, C. F. Eagen, Opt. Lett. 4, 236-238 (1979).
26. Krishanu Ray, et al., Appl. Phys. Lett., **90**, 251116/1-3 (2007).
27. I. Avrutsky, J Opt. Soc. Am. A **20**, 548-556 (2003) and references therein.
28. K. Selanger, A. J. Falnes, T. Sikkeland, *J. Phys. Chem.* 81, 1960-1963 (1977).

Figure 1.

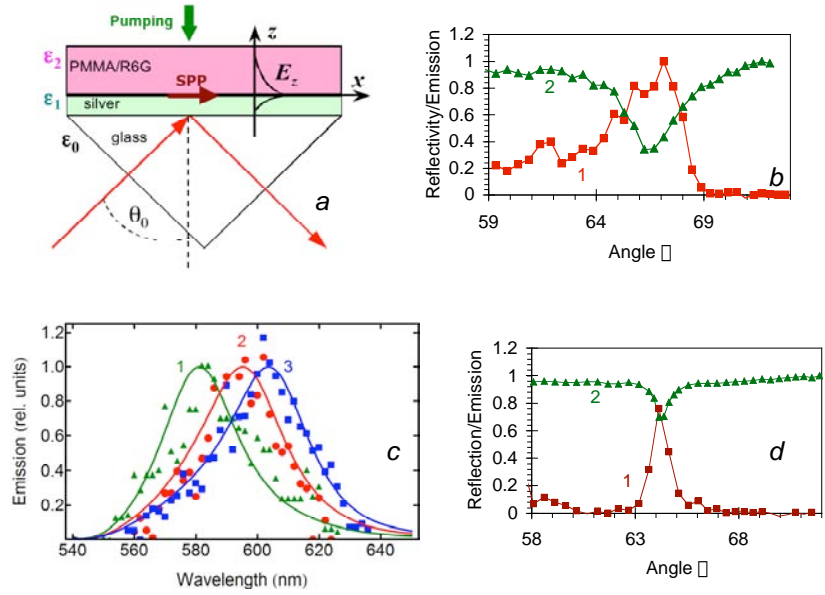

Figure 1. (a) Experimental sample; excitation and decoupling of SPP. (b) Angular profile of emission originating from decoupling of scattering-induced SPPs (trace 1) and the angular profile of reflectivity (trace 2); λ=575 nm. (c) Spectra of SPP spontaneous emission decoupled at different angles $\theta$; characters – experiment; solid lines – transfer matrix simulations; triangles and trace 1 – $\theta$=67.17°, circles and trace 2 – $\theta$=66.14°, squares and trace 3 – $\theta$=65.62°. The thickness of the silver film is 57 nm. (d) Angular profile of emission originating from decoupling of scattering-induced SPPs (trace 1) and the angular profile of reflectivity (trace 2). λ=632.8 nm.

Figure 2.

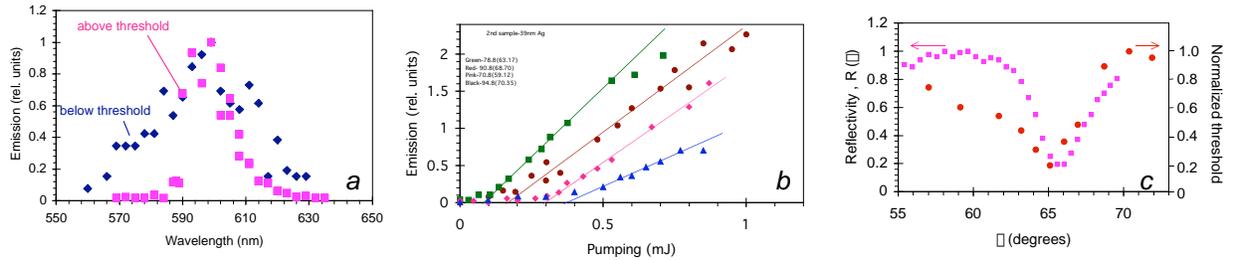

Figure. 2. (a) Spectra of the SPP emission recorded at $\theta=68.7°$ at low pumping density 10.9 mJ/cm$^2$ (diamonds) and high pumping density 81.9 mJ/cm$^2$ (squares). (b) Input-output curves of SPP emission recorded at different angles $\theta$. The thickness of the silver film is 39 nm. The diameter of the pumped spot is 2.16 mm. Squares – $\theta = 63.17°$, circles – $\theta = 68.70°$, diamonds – $\theta = 59.12°$, triangles – $\theta = 70.35°$. Resonance angle – $\theta_0 = 65.56°$. (c). Dependence of the SPP stimulated emission threshold *vs* $\theta$ (circles) and the reflectivity profile (squares) measured at $\lambda \approx 604$ nm.

Figure 3.

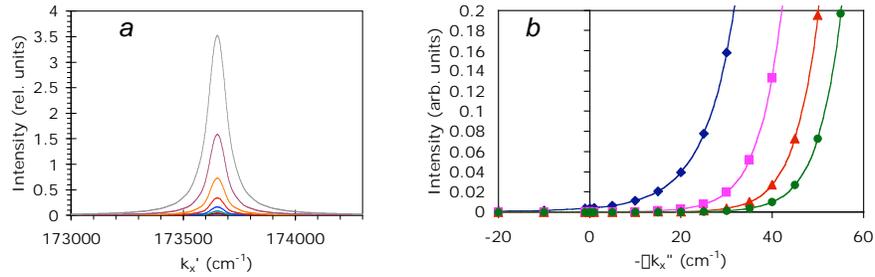

Figure 3. (a) Calculated SPP spectra in the wavenumber domain. The values of gain in the SPP system (from high to low) correspond to $k_x''$=-50, -45, ..-10, -5, -1, 0 cm$^{-1}$. $\varepsilon_0 = n_0^2 = 3.18$, $\varepsilon_1 = -13.6+i0.75$, $\varepsilon_2 = 2.25$; $\lambda = 594.1$ nm). The thickness of the silver film 39 nm. (b) Input-output curves of SPP emission calculated above the threshold at $k_x'$ = 173,660 cm$^{-1}$ (close to the resonance) – diamonds, 173,740 cm$^{-1}$ – squares, 173,860 cm$^{-1}$ – triangles, and 174,000 cm$^{-1}$ – circles.